\documentstyle[12pt]{article}
\begin{document}
\title{ Gribov Copies and Smeared Correlation Functions in Lattice QCD
}
\author{Maria Luigia Paciello$^{(1)}$, Claudio Parrinello$^{(2)}$,\\
        Silvano Petrarca$^{(1,3)}$, Bruno Taglienti$^{(1)}$,\\
        Anastassios Vladikas$^{(1)}$
\\[1.5em]
$^{(1)}$:INFN, Sezione di Roma {\it La Sapienza}\\
P.le A. Moro 2, 00185 Roma, Italy\\[0.5em]
$^{(2)}$: Physics Department,\\
New York University\\
4 Washington Place,\\
New York, NY 10003, USA,\\
and Physics Department,\\
Brookhaven National Laboratory,\\
Upton, NY 11973, USA\\[0.5em]
$^{(3)}$: Dipartimento di Fisica,\\
Universit\`a di Roma {\it La Sapienza},\\
P.le A. Moro 2, 00185 Roma, Italy\\[0.5em]
}
\maketitle
\newpage

\begin{abstract}
We study the influence of Gribov copies in the Coulomb
gauge on the smeared hadronic correlation functions that are involved in the
determination of the $B$ meson decay constant.
We find that the residual gauge freedom associated to Gribov copies
induces observable noise effects, though at the level of numerical accuracy
of our simulation these effects are not relevant to the
final determination of $f_B$.
Our results indicate that such effects may become important on
 bigger lattices.
\end{abstract}
\vfill
\newpage

In the last few years numerical studies of lattice gauge-fixing
ambiguities have been performed by several groups [1-4].
Such studies are interesting for many reasons. On the theoretical side,
the existence of these ambiguities parallels an analogous problem in the
continuum formulation of nonabelian gauge theories [5-10],
and the fact that
these nonperturbative features of the theory are reproduced both by continuum
and lattice regularized models is in itself reassuring.
Moreover, in the analytical study of the continuum limit of lattice gauge
theories\cite{Zwa,Gli} gauge-fixing is inescapable and the study of Gribov
ambiguities turns out to be an important point in
such a program. Finally, we mention that gauge dependent quark and
gluon matrix elements may be used to derive renormalization conditions, as
recently pointed out in \cite{Guidomaia},
and gauge-fixing ambiguities may play a role in
this kind of application.

{}From a more practical point of view, we observe that our current
understanding
of QCD phenomenology on the lattice very often proceeds through Monte Carlo
simulations involving gauge dependent operators, like hadron wavefunctions
and, in particular, ``smeared" sources for hadronic correlation functions
\cite{Parisi,Shar}.
In such cases, if
the gauge condition  we implement numerically (e.g. the
Coulomb gauge) does not correspond to a complete gauge-fixing,
in the sense that the gauge-fixing algorithm may converge randomly to any
configuration in a set of Gribov copies, then
the value of the operators will depend on which copy gets
selected by the algorithm. As we will discuss below,
such residual gauge freedom acts as a source
of statistical noise in the Monte Carlo average of those physical quantities
that are extracted from gauge dependent quantities.

The physics of the $B$ meson on the lattice is a suitable sector for an
investigation of the above issue.
The lattice calculation of the decay constant $f_B$
is performed in the static approximation,
in which the heavy constituent quark has infinite mass and, consequently,
the heavy quark propagator is given by a product of time-like links\cite{Eic}.

Given the local, gauge invariant ``heavy-light" axial current
\begin{equation}
A_{\mu}^{L} (\vec{x}, t) \equiv \bar{b}
(\vec{x}, t) \gamma_{\mu} \gamma_{5} q(\vec{x}, t)
\label{eq:currl}
\end{equation}
(where $b, q$ indicate field operators associated to a heavy and a light quark
respectively) one is interested in the asymptotic behaviour in time of the
2-point correlation function at zero spatial momentum:
\begin{equation}
C^{L L} (t) \equiv \sum_{\vec{x}} < 0 | T \{ A^{L}_{4} (\vec{x}, t) \ A_{4}^{L}
(0,1) \} | 0 >  \approx Z^{L} Z^{L} \ e^{- {\cal{E}}_B t}
\label{eq:correl}
\end{equation}
In principle $f_B$ can be obtained from $Z^L$, but in actual simulations
it has not been possible to isolate the lightest pseudoscalar state in the
above correlation function, thus no direct evaluation of $Z^L$ from $C^{LL}$
can be obtained\cite{bou}.
This problem has been overcome by evaluating on the lattice
the correlation
function associated with the ``smeared", gauge dependent, axial current
\begin{equation}
A_{\mu}^{S} (\vec{x}, t) \equiv {1 \over n^{3}} \ \sum_{i} \ \{\bar{b}
(\vec{x_{i}}, t) \gamma_{\mu} \gamma_{5} q(\vec{x}, t) \}
\label{eq:currs}
\end{equation}
The sum runs over $n^{3}$ lattice points in a 3-dimensional spatial cube
centred on $(\vec{x}, t)$, $n$ being an odd integer.
The 2-point function $C^{S S} (t)$ of the extended current (\ref{eq:currs}),
defined analogously to (\ref{eq:correl}), and the mixed 2-point function
$C^{S L} (t)$ are characterized by a much better numerical signal. Therefore
their (asymptotic) large $t$ behaviour
\begin{equation}
C^{S S} (t) \approx Z^{S} Z^{S} \ e^{- {\cal{E}}_B t}
\,\, {\rm{and}} \,\,
C^{S L} (t) \approx Z^{S} Z^{L} \ e^{- {\cal{E}}_B t}
\end{equation}
makes it possible, by taking ratios of these gauge dependent correlation
functions, to extract the gauge invariant $Z^{L}$\cite{guidobello}.

Such calculations are typically performed in the lattice Coulomb gauge.
In fact, it is well known that if one tries the numerical evaluation of
$C^{S S}$ and $C^{S L}$  without previously fixing the gauge, one experiences
large fluctuations induced by the gauge freedom, with no detectable
signal.
It is therefore conceivable that the residual gauge freedom associated to
an incomplete gauge fixing prescription (as we will see the lattice Coulomb
gauge, containing a large number of Gribov copies, is a good example) would
still induce unwanted fluctuations in the gauge dependent correlation functions
$C^{SL}$ and $C^{SS}$ and in their ratios, thus contributing to the overall
error bars for $Z^L$ and $f_B$.

The aim of this letter is to test numerically this possibility  by evaluating
$C^{S L}$ and $C^{S S}$ in the Coulomb gauge but on different Gribov
copies of the same thermalised configuration. We will follow closely
the procedures adopted in ref. \cite{guidobello}.
Before focusing on this, it is necessary to discuss the method that we
have adopted for the generation of such copies.

We consider a lattice with spatial volume $V_s$ and $T$ sites in the time
direction.
The standard lattice implementation of the Coulomb gauge
\cite{wilson,Davies}
consists of taking a thermalised link configuration $ \{ U \}$
and iteratively applying gauge transformations to it in order to
minimize the quantity
\begin{equation}
F [U^g] \equiv \frac{1}{T} \sum_{t=1}^{T}  f [U^g] (t)
\label{eq:bigf}
\end{equation}
where
\begin{equation}
f [U^g] (t) \equiv - \frac{1}{V_s}
Re \ Tr  \sum_{i=1}^{3} \sum_{\vec{x}} U_{i}^{g}
(\vec{x}, t)
\label{eq:smallf}
\end{equation}
and the gauge field transforms as $U_{i}^{g} (\vec{x}, t) =
g(\vec{x}, t) U_{i} (\vec{x}, t) g^{\dagger}(\vec{x}+\vec{i}, t)$.
The lattice Coulomb  gauge condition is automatically satisfied
when the transformed lattice $\{ U^g \}$ is such that $F[U^g]$, considered
as a function of the gauge transformations $g$, attains a local minimum.

We emphasize that by deciding to implement the Coulomb gauge through the
minimization of $F[U^g]$ one has already eliminated some of the lattice
Gribov copies of such a gauge, since any stationary point of $F[U^g]$
(including saddle points and local maxima) would be a configuration satisfying
the gauge condition, hence a Gribov copy. For this reason the minimization of
$F$ implements a gauge containing additional constraints with respect to the
standard Coulomb gauge. This corresponds to the original proposal made by
Gribov\cite{Gribov} for the quantization of the theory
 in the continuum, but it still turns out in both lattice and continuum models
that one has not eliminated all the gauge freedom, since in general
$F[U^g]$ as a function of $g$ has many local
minima\cite{DAZW,mari}. These are
the Gribov copies we will be dealing with.

A set of Gribov copies for each Monte Carlo thermalised configuration
can be generated in several ways. One may perform random gauge
transformations on the thermalised configuration before fixing the gauge,
or change the route of updating the lattice sites when gauge fixing, or
even vary a relevant parameter of the gauge fixing algorithm, as in the
overrelaxation procedure\cite{MandOlg,over}.

Eqs. (\ref{eq:bigf}) and (\ref{eq:smallf}) imply that, given a starting link
configuration, the Coulomb gauge condition is implemented independently on
each timeslice.
This is because the links in the time direction which connect adjacent
timeslices do not appear in the definitions of $F[U^g]$ and $f[U^g](t)$.
In other words, each timeslice of a given configuration is endowed with its
own scenario of Gribov copies, so that the pattern of their occurrence
in the Coulomb gauge is richer than in the Landau gauge, in which
they are defined globally on the whole lattice.

Consider then a gauge transformation $\tilde{g} $, that transforms the
original Monte Carlo configuration $\{ U \}$ into the Coulomb gauge-fixed
one $\{ U^{\tilde{g}} \}$.
Since Coulomb gauge-fixing is an independent process on each timeslice, it
is convenient to think of the set of matrices $\tilde{g}$ as a union of
subsets $\tilde{g} \equiv \bigcup_{t} \ \tilde{g}_{t}$, where the
$\tilde{g}_{t}$ implement the gauge on each timeslice\cite{Deforca}.
It is clear that if $\tilde{g}^{1} \equiv \ \bigcup_{t} \tilde{g}^{1}_{t}, \
\ \tilde{g}^{2} \equiv \bigcup_{t} \ \tilde{g}^{2}_{t}, \ \ \ldots, \
\tilde{g}^{N} \equiv \bigcup_{t} \ \tilde{g}^{N}_{t}$ are $N$ distinct
gauge transformations, that rotate the configuration $\{ U \}$ into
different realizations of the Coulomb gauge, then any combination
$\tilde{g}^{comb} \equiv \bigcup_{t} \ \tilde{g}^{i}_{t}$, where for each value
of $t$ the index $i$ can take any value between 1 and $N$, is a lattice gauge
transformation that implements a new distinct Coulomb gauge. In other words, a
Coulomb gauge transformation for the entire lattice can be built up from any
of the $T^{N}$ combinations of partial, fixed-time Coulomb gauges.
This property allows us to single out of the $T^{N}$ possible choices two
"special" realizations of the Coulomb gauge as follows: for each timeslice
$t$ of the configuration $\{ U \}$ we select from the set of fixed time
Coulomb gauges the transformation $\tilde{g}^{min}_{t} $ such that
$f [U^{\tilde{g}^{min}_{t}}] (t)$ takes the smallest value.
In such a way one can define ${g}^{min} \equiv \bigcup_{t} \
\tilde{g}^{min}_{t} $ as the Coulomb gauge transformation such that
$F[U^{g^{min}}]$ takes the smallest possible value.
Analogously, one can build up from our set of fixed time gauge transformations
${g}^{max}$, defined as the gauge transformation such that $F[U^{g^{max}}]$
takes the largest possible value.

In the following we will analyze the variation of the gauge dependent
operator
$C^{SS}(t)$ when evaluated on $\{ U^{g^{max}} \}$ and $ \{ U^{g^{min}} \}$,
comparing it to the corresponding variation of $f(t)$. In this way we can
estimate the order of magnitude of the fluctuations of $C^{SS}(t)$ induced
in an actual simulation by the residual gauge freedom associated to the Gribov
copies.
Such an estimate is based on the assumption that the configurations
$\{ U^{g^{max}} \}$ and $\{ U^{g^{min}} \}$, that have been constructed so
that they maximize $ \Delta [F] \equiv | F[U^{g^{max}}] - F[U^{g^{min}}] |$,
also maximize the variation of $C^{SS}(t)$. In some sense, we are assuming that
$F$ is a reliable measure of the ``distance" between Gribov copies.
In any case, our estimate should provide a lower bound for the magnitude of
the effect.
We have also obtained numerical results for $C^{SL}$, that are qualitatively
of the same type as those for $C^{SS}$ but characterized by a worse signal,
so that we prefer to focus the discussion on $C^{SS}$.

We have considered five $SU(3)$ lattice configurations generated on a
lattice of size $10^3 \times 20$ at $\beta = 6.0$.
The lattice has been doubled in the $x$ and $t$ directions before
evaluating the quenched light quark propagator for Wilson fermions
at $K=0.1515$. For these lattice parameters, the best smearing size is
$n=7$, according to \cite{guidobello}.
We have fixed the gauge to an accuracy ${\Delta [f] \over {| f |}
} \le 10^{-10}$
for each time slice; this eliminates the possibility of misinterpreting
fluctuations of $f$ related to the poor level of gauge fixing as the
evidence of distinct Gribov copies.

In the Tables (\ref{tab1}) and (\ref{tab2}) we report numerical results (half
lattice) from two of the five configurations that we have analyzed.
As can be seen from the second column of the tables, up to 6 distinct
values of $f(t)$, i.e. 6 different Gribov copies, have been obtained
on some timeslices. On other timeslices it has not been possible to
obtain as many as 6 distinct values, not even after generating 300 random
configurations, and in some cases no configurations with different
values of $f(t)$ were found. This last feature takes place in all 5
thermalised configurations, but obviously not on the same timeslices.

The two tables show different behaviors of $C^{SS}(t)$: in Table (\ref{tab1})
we see that $C^{SS}(t)$ fluctuates only when the value of $f(t)$ fluctuates,
while in Table (\ref{tab2}) $C^{SS}(t)$ always changes, even at those times
when $f(t)$ does not. The difference is due to the fact that, contrary to the
case of Table (\ref{tab1}), in Table (\ref{tab2}) $f$ fluctuates on the first
timeslice, so that $C^{SS}(t)$ is expected to change, since one of the smeared
sources is always located on the first timeslice.

In particular, the results  of Table (\ref{tab1}) confirm the existence of a
correspondence
between the fluctuations of $f(t)$ and those of the
smeared
hadronic sources at the same $t$, that generate the fluctuations
that we measure on
$C^{SS}(t)$. This justifies {\it a posteriori} our decision to probe the
fluctuations of hadronic correlation functions by evaluating them on those
configurations that maximize the fluctuations of $f(t)$.

In all cases examined we found, as expected, that $C^{SS}$ fluctuates
percentually much more than $f$. In particular, by inspection of
Table 1 it turns out that while $ {\Delta [f] \over {| f |}}$
is always less than
$1 \%$, the percentual variation of $C^{SS}(t)$ increases with $t$,
i.e. with the separation of the smeared sources, starting from less
than $1\%$ for $t=1$, going to more than $10\%$ at $t=9$ and being
$\approx 50\%$ or more for $t \geq 15$.

This is related to the fact that  $f(t)$ is the average on the timeslice
$t$ of a local function of the link variables, hence is sensitive to the
"noise" effect induced by the Gribov copies on that single timeslice;
on the contrary $C^{SS}$ is a 2-point correlation function
of fermionic operators and picks up a "noise" effect from the Gribov copies
on both the timeslices where the smeared
sources are located. Moreover, while the signal that we want to extract from
$C^{SS}(t)$ dies out exponentially with $t$, the "gauge noise" effect is
expected not to depend on the separation of the sources, thus becoming more
and more important percentually at large separations.

However, the gauge noise does not produce on our present lattice a
significant difference in the physics that we are considering, since the
signal is very noisy at large times, thus only the time interval
$t=5-10$ may be used for the determination of $Z^S$ and $Z^L$,
as in ref.\cite{guidobello}.
In such interval the gauge noise effect is not the dominant contribution
to the numerical uncertainty contained in the procedure for the determination
of $f_B$, that is mostly affected by errors due to the fitting of the
correlation functions.

Nonetheless, it is worth noting that the measurement of $f_B$ is not spoiled by
the gauge noise also because in the range of interest for $t$, that is between
5 and 10, in many cases we have not been able to find copies.
This fact reduces drastically our capability to fully evaluate the gauge noise
effect expected on this lattice.

On the basis of our results, we conclude that the residual gauge freedom
in the standard lattice implementation of the Coulomb gauge indeed
shows up as a noise effect on smeared hadronic correlation functions.

While the gauge noise is not too important in the framework of the actual
measurement that we have discussed, since this is plagued by many sources of
error, on the other hand, for the reasons that we have sketched above,
such an effect might sensibly affect measurements on bigger lattices.
In fact, increasing the lattice size would typically reduce the standard
statistical noise, allowing in principle to perform measurements from large
time separations, but then the gauge noise effect (that is not expected to
disappear when approaching the continuum limit)
may become a major source of fluctuations and provide a relevant contribution
to the error bars on the physical quantities evaluated from smeared
correlations.
We plan to further investigate the subject by increasing
the lattice size and the statistics and also by considering other smeared
operators.
\bigskip

We thank V. Lubicz, G. Martinelli and M. Testa for helpful discussions.
We also thank G. Martinelli for allowing us to use an analysis program.

C.P. acknowledges financial support from C.N.R. and I.N.F.N.

\newpage
\begin{table}
\centering
\begin{tabular}{|c|c|c|c|c|c|}\hline
$t$ &N& $f[U^{g^{min}}]$ & $f[U^{g^{max}}]$ &
 $C^{SS}[U^{g^{min}}]$ &
$C^{SS}[U^{g^{max}}]$ \\ \hline
1  &1&  -2.640147  & -2.640147 &  0.1152 $\cdot$ $10^{-1}$ &  0.1152 $\cdot$
$10^{-1}$ \\
2  &2&  -2.625052  & -2.623131 &  0.5371 $\cdot$ $10^{-2}$ &  0.5149 $\cdot$
$10^{-2}$ \\
3  &1&  -2.641978  & -2.641978 &  0.2548 $\cdot$ $10^{-2}$ &  0.2548 $\cdot$
$10^{-2}$ \\
4  &2&  -2.640007  & -2.639101 &  0.1226 $\cdot$ $10^{-2}$ &  0.1242 $\cdot$
$10^{-2}$ \\
5  &2&  -2.628902  & -2.625713 &  0.5600 $\cdot$ $10^{-3}$ &  0.4558 $\cdot$
$10^{-3}$ \\
6  &3&  -2.623502  & -2.621745 &  0.3147 $\cdot$ $10^{-3}$ &  0.2693 $\cdot$
$10^{-3}$ \\
7  &1&  -2.640893  & -2.640893 &  0.1970 $\cdot$ $10^{-3}$ &  0.1970 $\cdot$
$10^{-3}$ \\
8  &1&  -2.635022  & -2.635022 &  0.1056 $\cdot$ $10^{-3}$ &  0.1056 $\cdot$
$10^{-3}$ \\
9  &2&  -2.637537  & -2.635372 &  0.5881 $\cdot$ $10^{-4}$ &  0.6239 $\cdot$
$10^{-4}$ \\
10 &1&  -2.629200  & -2.629200 &  0.3266 $\cdot$ $10^{-4}$ &  0.3266 $\cdot$
$10^{-4}$ \\
11 &4&  -2.614060  & -2.608795 &  0.1392 $\cdot$ $10^{-4}$ &  0.7761 $\cdot$
$10^{-5}$ \\
12 &5&  -2.620060  & -2.611935 &  0.7838 $\cdot$ $10^{-5}$ &  0.9910 $\cdot$
$10^{-5}$ \\
13 &6&  -2.627113  & -2.624630 &  0.4899 $\cdot$ $10^{-5}$ &  0.5967 $\cdot$
$10^{-5}$ \\
14 &5&  -2.621742  & -2.616905 &  0.2030 $\cdot$ $10^{-5}$ &  0.1667 $\cdot$
$10^{-5}$ \\
15 &5&  -2.620936  & -2.615650 &  0.4411 $\cdot$ $10^{-6}$ &  0.1825 $\cdot$
$10^{-5}$ \\
16 &4&  -2.630487  & -2.622896 &  0.2994 $\cdot$ $10^{-6}$ &  0.3970 $\cdot$
$10^{-6}$ \\
17 &4&  -2.621350  & -2.619444 &  0.1468 $\cdot$ $10^{-5}$ &  0.9535 $\cdot$
$10^{-6}$ \\
18 &2&  -2.618932  & -2.618043 &  0.4467 $\cdot$ $10^{-6}$ &  0.4002 $\cdot$
$10^{-6}$ \\
19 &2&  -2.634834  & -2.634378 &  0.4774 $\cdot$ $10^{-6}$ &  0.3862 $\cdot$
$10^{-6}$ \\
20 &1&  -2.627090  & -2.627090 &  0.2009 $\cdot$ $10^{-6}$ &  0.2009 $\cdot$
$10^{-6}$ \\
\hline
\end{tabular}
\caption{
The values per timeslice of the
gauge dependent quantities $f(t)$ and $C^{SS}(t)$ for the two
Gribov copies $U^{g^{min}}$ and $U^{g^{max}}$.
$N$ is the number of distinct values for $f(t)$ we found
on each timeslice.
}
\label{tab1}
\end{table}

\newpage
\begin{table}
\centering
\begin{tabular}{|c|c|c|c|c|c|}\hline
$t$ &N& $f[U^{g^{min}}]$ & $f[U^{g^{max}}]$ &
 $C^{SS}[U^{g^{min}}]$ &
$C^{SS}[U^{g^{max}}]$ \\ \hline
1  &3&  -2.621564  & -2.619602 &  0.1089 $\cdot$ $10^{-1}$ &  0.1092 $\cdot$
$10^{-1}$ \\
2  &1&  -2.631229  & -2.631229 &  0.3748 $\cdot$ $10^{-2}$ &  0.4008 $\cdot$
$10^{-2}$ \\
3  &3&  -2.622969  & -2.621627 &  0.1805 $\cdot$ $10^{-2}$ &  0.1796 $\cdot$
$10^{-2}$ \\
4  &1&  -2.627733  & -2.627733 &  0.7258 $\cdot$ $10^{-3}$ &  0.7160 $\cdot$
$10^{-3}$ \\
5  &1&  -2.633529  & -2.633529 &  0.4288 $\cdot$ $10^{-3}$ &  0.4195 $\cdot$
$10^{-3}$ \\
6  &2&  -2.638506  & -2.636724 &  0.2006 $\cdot$ $10^{-3}$ &  0.1897 $\cdot$
$10^{-3}$ \\
7  &2&  -2.639759  & -2.637878 &  0.9609 $\cdot$ $10^{-4}$ &  0.8535 $\cdot$
$10^{-4}$ \\
8  &2&  -2.627650  & -2.625860 &  0.5094 $\cdot$ $10^{-4}$ &  0.5585 $\cdot$
$10^{-4}$ \\
9  &1&  -2.637258  & -2.637258 &  0.2425 $\cdot$ $10^{-4}$ &  0.2118 $\cdot$
$10^{-4}$ \\
10 &4&  -2.627042  & -2.623546 &  0.1151 $\cdot$ $10^{-4}$ &  0.1205 $\cdot$
$10^{-4}$ \\
11 &5&  -2.620013  & -2.616689 &  0.6815 $\cdot$ $10^{-5}$ &  0.4685 $\cdot$
$10^{-5}$ \\
12 &3&  -2.620194  & -2.615002 &  0.1962 $\cdot$ $10^{-5}$ &  0.2173 $\cdot$
$10^{-5}$ \\
13 &6&  -2.611956  & -2.610578 & -0.4144 $\cdot$ $10^{-6}$ & -0.5794 $\cdot$
$10^{-6}$ \\
14 &3&  -2.626609  & -2.625656 & -0.1033 $\cdot$ $10^{-5}$ &  0.3282 $\cdot$
$10^{-6}$ \\
15 &2&  -2.637120  & -2.629574 &  0.5860 $\cdot$ $10^{-6}$ &  0.1294 $\cdot$
$10^{-5}$ \\
16 &3&  -2.633638  & -2.628927 & -0.8613 $\cdot$ $10^{-7}$ &  0.9807 $\cdot$
$10^{-6}$ \\
17 &2&  -2.629207  & -2.628761 & -0.9640 $\cdot$ $10^{-6}$ & -0.3911 $\cdot$
$10^{-6}$ \\
18 &2&  -2.638059  & -2.634357 &  0.4195 $\cdot$ $10^{-6}$ &  0.6314 $\cdot$
$10^{-6}$ \\
19 &1&  -2.635334  & -2.635334 &  0.2419 $\cdot$ $10^{-6}$ &  0.4509 $\cdot$
$10^{-6}$ \\
20 &2&  -2.618391  & -2.616922 &  0.1306 $\cdot$ $10^{-6}$ &  0.4818 $\cdot$
$10^{-6}$ \\
\hline
\end{tabular}
\caption{Same as in Table 1 for a different thermalised configuration.}
\label{tab2}
\end{table}
\end{document}